\documentclass[12pt,preprint]{aastex}
\def\HST{{\it HST}}

\shorttitle{M31 RV}
\shortauthors{Bond \& Siegel}

\begin{document}

% \title{HST Limits on the Nature of M31-RV\footnote{The standard HST footnote}}

\title{\emph{Hubble Space Telescope} Imaging of the Outburst Site of
M31~RV\altaffilmark{1}}

\author{Howard E. Bond\altaffilmark{2} and Michael H. Siegel\altaffilmark{3} }

\affil{Space Telescope Science Institute, 3700 San Martin Drive, Baltimore, MD
21218; bond@stsci.edu, siegel@astro.as.utexas.edu}

\altaffiltext{1}
{Based in part on observations made with the NASA/ESA {\it Hubble Space
Telescope}, obtained from the data archive at the Space Telescope Science
Institute. STScI is operated by the Association of Universities for Research in
Astronomy, Inc., under NASA contract NAS5-26555.}

\altaffiltext{2}
{Visiting astronomer, Kitt Peak National Observatory, which is operated by the
Association of Universities for Research in Astronomy, Inc., under cooperative
agreement with the National Science Foundation.}

\altaffiltext{3}
{Current address: University of Texas -- McDonald Observatory, Austin, TX
78712}

% version date May 4, 2004

\clearpage

\begin{abstract}

M31 RV is a luminous red variable star that appeared for several months in the
bulge of M31 during 1988. Unlike classical novae, M31 RV was cool throughout its
outburst. Interest in this object has revived recently because of its strong
resemblance to V838~Mon, a luminous Galactic variable star that appeared in 2002
and is illuminating a spectacular light echo, and has evolved to ever-cooler
surface temperatures. V4332~Sgr is a third object which was also a red
supergiant throughout its eruption.

We have examined archival {\it Hubble Space Telescope\/} ({\it HST\/}) images of
the site of M31~RV, obtained fortuitously in 1999 with the  WFPC2 camera in
parallel mode during spectroscopic observations of the nucleus of M31. We
located the site of M31~RV in the \HST\/ frames precisely through astrometric
registration with ground-based CCD images, including several taken during the
outburst.  No light echo is seen at the M31~RV site, implying that M31~RV is not
surrounded by circumstellar (or interstellar) dust similar to that around
V838~Mon, or that its extent is less than $\sim$1.7~pc.

The stellar population at the outburst site consists purely of old red giants;
there is no young population, such as seen around V838~Mon. There are no stars
of unusual color at the site, suggesting that M31~RV had faded below \HST\/
detectability in the 11~years since outburst, that it is an unresolved companion
of one of the red giants in the field, or that it {\it is\/} one of the red
giants. We suggest future observations that may help decide among these
possibilities.

\end{abstract}

\keywords{ stars: evolution---novae, cataclysmic variables---stars: individual
(M31 RV, V838 Mon, V4332 Sgr)---stars: variables: other}

\clearpage

\section{Introduction}

In mid-1988, an unusual stellar outburst in the nuclear bulge of the Andromeda
Galaxy, M31, was discovered independently by Rich et~al.\ (1989), Bryan \& Royer
(1991), and Tomaney \& Shafter (1992).  Although similar in luminosity to a
classical nova, the object was cool and red throughout its eruption. Its
behavior was thus completely different from that of a classical nova, in which
an extremely hot and blue remnant is quickly revealed as the ejected envelope
expands and becomes optically thin.  This remarkable object has been called the
``M31 red variable," or ``M31 RV.''

M31 RV's optical outburst light curve has been assembled from published
observations plus their own analyses of archival plate material by Sharov (1990,
1993) and more recently by Boschi \& Munari (2004).  Although the rise to
maximum was not well observed, M31~RV was brighter than 18.5 $B$ magnitude
($M_B\lesssim-6$) for at least 80 days, but then rapidly declined to
invisibility. Inspection of archival plates shows that the 1988 outburst was the
only one in the past half century (Boschi \& Munari 2004), the claim of a
previous eruption in 1968 (Sharov 1990) having subsequently been withdrawn
(Sharov 1993).

Spectroscopic observations of M31 RV near maximum and during the subsequent
decline showed a spectrum resembling that of an M0 supergiant  (Rich et al.\
1989), gradually evolving toward M5 and then late M as the outburst proceeded
(Mould et al.\ 1990). At maximum brightness, M31~RV was one of the most luminous
stars in the Local Group, at a bolometric absolute magnitude of $M_{\rm
bol}\simeq-10$ (Rich et al.\ 1989). Unfortunately, M31~RV was not well observed
during its outburst, and little else is known about this unusual event.

Interest in M31 RV has revived recently because of its striking resemblance to
V838 Monocerotis. V838~Mon, a previously unknown Galactic variable star, erupted
in 2002 January and reached a peak luminosity similar to that of M31 RV (Bond et
al.\ 2003; Munari et al.\ 2005; and references therein).  Its spectrum evolved
rapidly from type K to a very cool M and then L type, accompanied by formation
of a dense circumstellar dust envelope (Banerjee \& Ashok 2002; Evans et al.\
2003; Lynch et al.\ 2004; Rushton et al.\ 2005).  The outburst of V838 Mon was
followed by the appearance of a spectacular light echo (Henden, Munari, \&
Schwartz 2002; Munari et al.\ 2002; Bond et al.\ 2003; Crause et al.\ 2005),
imaged extensively by the {\it Hubble Space Telescope\/} (\HST\/) as well as
from the ground, which continues to evolve at the present time.

V4332 Sagittarii is a third object\footnote{Nova V1148~Sgr 1943 was reported to
have a late-type spectrum by Mayall (1949) based on three objective-prism
plates, making it a possible fourth member of the class, but nothing else is
known about this object. More speculatively, Kato (2003) has suggested that Nova
CK~Vul 1670 could have been a V838~Mon-like event, but of course nothing is
known about its outburst spectrum.} with similarities to M31~RV and V838~Mon. In
1994, V4332 Sgr had a nova-like outburst, during which it remained very cool
(Martini et al.\ 1999; Banerjee \& Ashok 2004; Tylenda et al.\ 2005 and
references therein). The absolute luminosity of V4332~Sgr is highly uncertain,
but if the star lies in the nuclear bulge of our own Galaxy, it was several
magnitudes less luminous than  M31~RV and V838~Mon at maximum light.

A number of explanations have been proposed for this new class of peculiar
outburst events. These include an outburst from a compact object in a red-giant
envelope (Mould et al.\ 1990); a hydrogen shell flash on the surface of an
accreting cold white dwarf in a short-period binary (Iben \& Tutukov 1992); a
thermonuclear event in a single star (Martini et al.\ 1999; Munari et al.\
2005); the merger of two main-sequence stars (Soker \& Tylenda 2003; Tylenda et
al.\ 2005); the accretion of planets by a giant star (Retter \& Marom 2003); and
a born-again red-giant event in a binary system (Lawlor 2005). Moreover, Yaron
et al.\ (2005) hint that a complete exploration of the parameter space for
classical novae might reveal models with properties similar to these objects.
The  discovery that V838~Mon has an unresolved B-type companion (Munari \&
Desidera 2002; Wagner \& Starrfield 2002), as well as belonging to a small
cluster containing several more B-type stars (Afsar \& Bond 2005), and that it
therefore may have arisen from a fairly massive progenitor star, may rule out
some of these scenarios, but a fully convincing explanation remains elusive.

The plethora of mutually exclusive scenarios shows that these objects pose a
significant challenge to our understanding of stellar physics. In an effort to
provide further information on M31~RV, we have examined archival images of the
site of the event obtained with \HST\/ some 11 years after the outburst.  The
aims of our investigation are to determine whether M31~RV produced a detectable
light echo, to characterize the stellar population surrounding the object, and,
if possible, to detect a remnant star.

\section{Archival \emph{HST} and Ground-Based Images}

M31~RV has never been targeted specifically by \HST, but the bulge of M31 has
been imaged in several unrelated programs. We searched the archive\footnote{The
\HST\/ data archive is available at http://archive.stsci.edu/hst} for images
that serendipitously cover the location of M31~RV, obtained with any of the
\HST\/ cameras: Wide Field and Planetary Camera (WF/PC), Wide Field Planetary
Camera~2 (WFPC2), Space Telescope Imaging Spectrograph (STIS), Near Infrared
Camera and Multiobject Spectrograph (NICMOS), and Advanced Camera for Surveys
(ACS)\null.  We found that no images of the site have been obtained with STIS or
NICMOS, nor with WF/PC (although one set of frames taken in 1992 July missed
M31~RV by only $1\farcs6$!). 

With WFPC2, there are two sets of archival observations that do include the site
of M31~RV\null.  One set, called hereafter the ``u58 series,'' was taken on 1999
July 23--24 in program GTO-8018 (PI: R.~F.\ Green); these WFPC2 images were taken
in parallel mode during STIS spectroscopy of the nucleus of M31, and thus the
inclusion of M31~RV in the WFPC2 field is purely fortuitous. The u58 series
consists of datasets u5850101r through u5850108r, taken through the ``$V$''
filter (F555W) (consisting of 8 dithered exposures of 600--1000~s), and datasets
u5850109m through u585020br, taken through the ``$I$'' filter (F814W)
(consisting of 13 exposures of 300--1000~s). In these images, M31~RV lies in the
high-resolution Planetary Camera (PC) chip.

An earlier set of six observations, which we call the ``u31 series,'' was taken
on 1995 December 5 in program GTO-6255 (PI: I.~King), which was also a STIS
spectroscopic program on the M31 nucleus, with the WFPC2 images again being
taken in parallel. The u31 series contains two 1300~s exposures through the
F300W ultraviolet filter (u31k0109t and u31k010bt), and four 2700~s exposures
through the F170W ultraviolet filter (u31k010ft through u31k010rt). In the u31
series, the site of M31~RV lies in one of the low-resolution Wide Field (WF)
chips. 

Finally, with the ACS, there are two \HST\/ observations fortuitously showing
the M31~RV site, both of them 2200~s exposures in the ``$B$'' band (F435W),
taken in program GO-10006 (PI: M.~Garcia) as part of a project on X-ray novae in
M31. Exposures were taken on 2003 December 12 (dataset j8vp02010) and 2004
October~2 (j8vp07010), the latter being unavailable to us at this writing due to
the proprietary \HST\/ data policy.

M31~RV lies in an extremely dense stellar field, which is well resolved in
\HST\/ images. It is therefore desirable to locate the site of the outburst
event as accurately as possible, as a preliminary to investigation of the \HST\/
frames.  We have done this using two sets of ground-based CCD images that
include the site of M31~RV, one of which shows the object during its outburst.

The first ground-based set was obtained by R.~Ciardullo on 1988 September 29
with the No.~1 0.9-m reflector at Kitt Peak National Observatory (KPNO), and was
kindly made available to us. These images were obtained fortuitously during the
eruption of M31~RV, as part of a search for classical novae in the bulge of M31,
and contain well-exposed images of M31~RV (which was  cataloged as ``Nova~36''
in Ciardullo et al.\ 1990).  There are two 900~s exposures taken through a
narrow-band H$\alpha$ filter, and one 420~s frame taken through a wider-band
continuum filter centered at 6091~\AA.

Because of the small field of view of the \HST\/ images, and the relative
shallowness of the 0.9-m frames, there are too few detectable stars seen in
common in both the \HST\/ and 0.9-m frames to allow them to be registered
astrometrically. We therefore also used a second, deeper set of ground-based
frames of the M31 bulge, which had been obtained by H.E.B.\ for a different
purpose with the KPNO 4-m Mayall telescope and its Mosaic camera on 1999
January~17, and which fortuitously cover the location of M31~RV\null. These
frames show the brightest stars seen in the \HST\/ images, while having a large
enough field to also show several stars visible in the 0.9-m frames. We chose
for further analysis two 30~s exposures taken in good seeing through a
Kron-Cousins $I$ filter.

\section{Astrometry}

\subsection{Absolute Astrometry of M31~RV}

We first used the three KPNO 0.9-m frames that show M31~RV in eruption to derive
absolute astrometry of the object. We identified 9 nearby field stars contained
in the NOMAD-1.0 astrometric catalog (Zacharias et al.\ 2004; optical
photographic survey plates in the bright M31 bulge are generally saturated, and
most of the NOMAD positions at this location are derived from stellar
coordinates in the 2MASS infrared catalog). These field stars were used to
establish an astrometric grid on each of the three frames, from which we
obtained a position for M31~RV accurate to about $\pm$$0\farcs04$ in each
coordinate, based on the rms scatter in the astrometric solutions and the
excellent agreement among our three CCD frames.

The absolute position of M31~RV is given in Table~1 along with, for comparison,
the positions derived by Munari, Henden, \& Boschi (2003) from two photographic
plates obtained with the Asiago Schmidt telescope during the outburst. (Munari
et al.\ also list several earlier, less-precise position measurements from the
literature.)  The agreement is satisfactory, given the smaller plate scale of
the Asiago material and the necessity to set up secondary astrometric standards
as part of their solutions.  This position lies $4\farcm7$ to the southeast of
the nucleus of M31, corresponding to a projected linear separation of 1.0~kpc.

\subsection{Astrometric Registration of the Ground-Based and \emph{HST} Images}

We then proceeded to locate the site of M31~RV on the \HST\/ frames. As noted
above, we cannot directly register the KPNO 0.9-m frames showing M31~RV with the
\HST\/ frames, since there are insufficient visible stars in common. 

We therefore used the KPNO 4-m frames in an intermediate step, as follows.
First, we selected the 0.9-m frame with the best seeing, and identified 9 stars
visible both on this frame and an image created by registering and combining the
two excellent 4-m frames. We then used the IRAF\footnote{IRAF is distributed by
the National Optical Astronomy Observatory, which is operated by the Association
of Universities for Research in Astronomy, Inc., under cooperative agreement
with the National Science Foundation.} routine {\it geomap\/} to compute the
geometric transformation to be applied to the 0.9-m image so as to register it
with the 4-m frame, and then the {\it geotran\/} routine to actually apply this
transformation. This allowed us to mark the location of M31~RV preciesly in the
4-m frame.

Next we applied the IRAF routine {\it wmosaic\/} to \HST\/ images in the u58
series so as to combine the four WFPC2 chips into single-image mosaics, and used
a similar geometric transformation from the 4-m image to locate the site of
M31~RV in the WFPC2 mosaic. Finally, we did a transformation from the WFPC2
mosaic to the PC chip to locate M31~RV in the latter.\footnote{The WFPC2 
reference image chosen was u5850103r, in which the derived location of M31~RV
lies in the PC chip at pixel coordinates $(x,y)=(549.6, 675.4)$.} A formal
error-propagation calculation indicates that the location in the PC chip is
accurate to $\pm$$0\farcs18$ = $\pm$3.9 PC pixels in the $x$ coordinate, and 
$\pm$$0\farcs27$ = $\pm$5.9 PC pixels in $y$.

\section{The Site of M31 RV}

\subsection{Visual Examination}

We prepared high-S/N, cosmic-ray rejected WFPC2 images of the M31 RV site by
registering and combining all of the u58 series $V$ and $I$ images. Figure~1
illustrates the location of M31~RV in the combined \HST\/ images. These are
$3''\times3''$ images centered on the derived location of the outburst.

The sheet of stars belonging to the bulge of M31 is well resolved in these deep
$V$ and $I$ frames. There are no obvious stars with unusual colors or
brightnesses at the M31~RV location. Detailed stellar photometry of the field
will be presented below. 

We also examined the ultraviolet (F300W and F170W) WFPC2 frames from the u31
series visually. These frames show very few stars, and there is nothing obvious
at the M31~RV site. The single ACS $B$-band image currently available in the
\HST\/ archive likewise shows no stars of unusual colors at the outburst site.

\subsection{Absence of Light Echo}

Although the WFPC2 frames in Figure~1 are considerably deeper than \HST\/ frames
that show the light echo around V838~Mon, no such feature is visible at the
location of M31~RV at the 1999.6 epoch.

The geometry of a light echo is simple (e.g., Bond et al.\ 2003 and references
therein): at a time $t$ after the outburst, the illuminated dust lies on the
paraboloid given by $z=x^2/2ct-ct/2$, where $x$ is the projected distance from
the star in the plane of the sky, $z$ is the distance from this plane along the
line of sight toward the Earth, and $c$ is the speed of light.

Figure~2 illustrates the geometry for M31~RV\null. It shows the light-echo
paraboloids at $t=2$~through 10~yr after the outburst, with a spacing of 2~yr,
and, as a darker line, the parabola at the time of the u58 series of \HST\/
images, 11.0~yr after the outburst. Also shown at the top is the conversion to
angular units, at the nominal 725~kpc distance of M31.  The dashed circle shows
a radius of 2~pc around the star, which is the approximate outer boundary of
illuminated dust currently being seen around V838~Mon (e.g., Bond et al.\ 2003;
Crause et al.\ 2005).

This figure demonstrates that, if M31~RV were surrounded by circumstellar dust
with an extent and density similar to that around V838~Mon, we should have seen
a light echo in the WFPC2 images taken in 1999.  The approximate diameter would
have been $\sim$$0\farcs8$, which would be readily resolved in Figure~1.

The absence of a light echo has two possible explanations: (1)~there is very
little circumstellar (or interstellar\footnote{Note that some authors (Tylenda
2004; Crause et al.\ 2005) have suggested that the light echo around V838~Mon
arises from ambient interstellar dust, rather than from material ejected from
the star.}) dust around M31~RV, or (2)~if there is circumstellar dust around
M31~RV similar in density to that around V838~Mon, Figure~2 shows that it
extends less than $\sim$1.7~pc from the star.  Figure~2 suggests that it might
be worthwhile to examine any existing high-resolution ground-based images of the
M31 bulge obtained around the early to mid-1990's, at which time any light echo
from circumstellar dust would have had maximum apparent radius.

\subsection{Stellar Photometry}

We performed stellar photometry on the u58 series images, using HSTphot (Dolphin
2000), a program that performs automated PSF-fitting photometry, accounts for
aperture corrections and charge-transfer effects, and transforms the
instrumental magnitudes to the Johnson-Kron-Cousins standard system. HSTphot
detected over 70,000 stellar objects in the WFPC2 field, 12,000 of which lie on
the PC chip that contains the location of M31~RV.

The quality of the photometry is excellent, with errors of less than 0.1~mag
down to $V\simeq25.5$ and $I\simeq24.7$ for isolated stars in the PC field. 
However, because of the extreme crowding in this field, the total star counts
roll over at  $V\simeq24$ and $I\simeq22$, with a steady decline over the next
two magnitudes. 

The color-magnitude diagram (CMD) for the entire PC chip is shown in the top
plot in Figure~3. The CMD is that of an old population of red giants, in
agreement with other studies of the M31 bulge (e.g., Jablonka et al.\ 1999;
Stephens et al.\ 2003; and references therein).  Apart from less than a dozen
blue objects (some of which could be background unresolved galaxies, or blue
stragglers in M31), there is no evidence for any significant young population at
this location in the M31 bulge.

The bottom plot in Figure 3 shows the CMD of stars within radii of 6 pixels
(star symbols), and radii of 6 to 18 pixels (filled circles); these radii
correspond to approximately 1$\sigma$ and 3$\sigma$ position errors, based on
the values given in \S3.2.  None of these stars have unusual colors; all of
them lie on the M31 bulge red-giant branch. 

\section{Discussion}

\subsection{The Stellar Population of M31 RV}

We have shown that M31~RV belongs to an old stellar population in the bulge of
M31. This is in striking contrast to V838~Mon, which has an unresolved B-type
companion, as well as belonging to a small cluster containing several additional
B-type stars.

The similarities of the light curves, spectral evolution, and peak luminosities
of M31~RV and V838~Mon strongly suggest a common outburst mechanism. If this is
the case, this mechanism must be one that can occur in stars belonging to both
young and old populations. Whether this constrains any of the scenarios
mentioned in \S1 is unclear. The lack of any young stars near M31~RV does
suggest, however, that the B-type companion of V838~Mon is a bystander that did
not play an essential role in the outburst.

\subsection{Searching for a Stellar Remnant}

All of the stars at the outburst site have the magnitudes and colors of ordinary
red giants in the M31 bulge. The absence of any conspicuous remnant star  has
three possible explanations: (a)~the object had faded below \HST\/ detectability
in the 11~years since outburst, either intrinsically or because of heavy dust
obscuration; (b)~the remnant is an unseen companion of (or its image is blended
with) one of the red giants in the field; or (c)~the remnant {\it is\/} one of
the red giants.

The post-outburst histories of V838 Mon and V4332~Sgr give us only modest
guidance. At the present time, V838~Mon remains enshrouded in dust, but it
continues to be luminous at long wavelengths. For example, in 2004 December it
had an apparent magnitude of $I\simeq10.5$ and an extremely red color of
$V-I\simeq4.7$ (Crause et al.\ 2005). Assuming $E(B-V)=0.9$ and a nominal
minimum distance of 6~kpc for V838~Mon (Munari, Desidera, \& Henden 2002; Bond
et al.\ 2003; Munari et al.\ 2005), and neglecting the small foreground
reddening for M31, the corresponding values if the V838~Mon of 2004 December
were located in M31 would be $I\lesssim19.4$ and $V-I\simeq3.4$. {\it Figures~1
and~3 show that there was no such bright, very red star at the location of
M31~RV in 1999.}\footnote{We also examined images from the 2MASS survey taken in
1997 October, which likewise show no red star at the M31~RV location; however,
V838~Mon as of 2004 December moved to the distance of M31 would have been
fainter than the 2MASS survey completeness limits.}

Of course, V838~Mon in 2004 December---less than 3 years after its outburst
maximum---is not the most suitable comparison object for M31~RV observed 11
years after its eruption. V4332~Sgr may provide a more apt comparison, since
more than 11 years have now elapsed since its eruption in early 1994. 

CCD photometry of V4332~Sgr is collected in Table~2; in addition to previously
published photometry at maximum light and in 2003, we include observations made
by us in 2004-5 at the 1.3-m SMARTS Consortium telescope at Cerro Tololo, which
have errors of about $\pm$0.05~mag. In recent years, V4332~Sgr has been
essentially constant near an apparent magnitude of $I\simeq15.1$ and a color of
$V-I\simeq2.5$. At maximum light (Martini et al.\ 1999, their Figure~2), the $I$
magnitude was about 7. The star has therefore declined from its maximum by about
8~mag at the present time. Note, however, that pre-outburst sky-survey images
show a star at the location of V4332~Sgr of approximately the same brightness as
the object seen at present. This fact, along with the constancy of the past few
years, may suggest that the 15th-mag object is a field star that is blended with
the variable (or is possibly a physical companion). In this case, the variable
has now declined by {\it more\/} than 8~mag. We note, though, that V4332~Sgr was
imaged with \HST\/ on 1997 November~3 (program SNAP-7386, PI: F.~Ringwald),
using WFPC2 with a narrow-band H$\alpha$ filter. The image, which lies in the PC
chip, shows no strong evidence for a resolved companion star, but the FWHM of
the stellar profile does appear marginally larger than for nearby field stars.

The $I$ magnitude of M31~RV at maximum was approximately 14 (see Boschi \&
Munari 2004).  If it had faded by only 8~mag by 1999.6, it would have been at
$I\simeq22$.  The reddening of V4332~Sgr is $E(B-V)\simeq0.32$ (Martini et al.\
1999), so its recent intrinsic color (see Table~2) is $(V-I)_0\simeq 2.1$.
Figure~3b shows that there {\it are\/} several stars with these characteristics
in the vicinity of M31~RV, although none of them are within 1$\sigma$ of the
site.

\subsection{Future Work}

We are left with the unsatisfying situation that there are several stars near
the site of M31~RV that could be the remnant, but which could also merely be 
normal field red giants.

There are several types of observations that could shed additional light on the
situation. These include (a)~a new \HST\/ observation, to see whether any of the
stars near the outburst site have faded since 1999; (b)~a grism observation with
\HST\/ to determine whether any of the stars near the site share the remarkable
very low-excitation emission-line spectrum now exhibited by V4332~Sgr (Banerjee
\& Ashok 2004; Tylenda et al.\ 2005); and (c)~near-infrared imaging with NICMOS
to see whether any of the stars show an IR excess similar to that currently
shown by the dust-obscured V838~Mon.

\acknowledgments

The authors thank Robin Ciardullo for providing a tape containing his 1988
images of M31~RV and Andrew Dolphin for assistance with the HSTphot program.
The SMARTS observations of V4332~Sgr were made by David Gonz\'alez and Juan
Espinoza. H.E.B.\ acknowledges many interesting discussions with Sumner
Starrfield.
This research has made use of the USNOFS Image and Catalogue Archive
operated by the United States Naval Observatory, Flagstaff Station
(http://www.nofs.navy.mil/data/fchpix/).
Support for this work was provided by NASA through grant number
GO-9587 from the Space Telescope Science Institute, which is operated by
AURA, Inc., under NASA contract NAS 5-26555.

\clearpage

% {\bf Figure Captions}

\begin{figure}
\begin{center}
\includegraphics[width=3in]{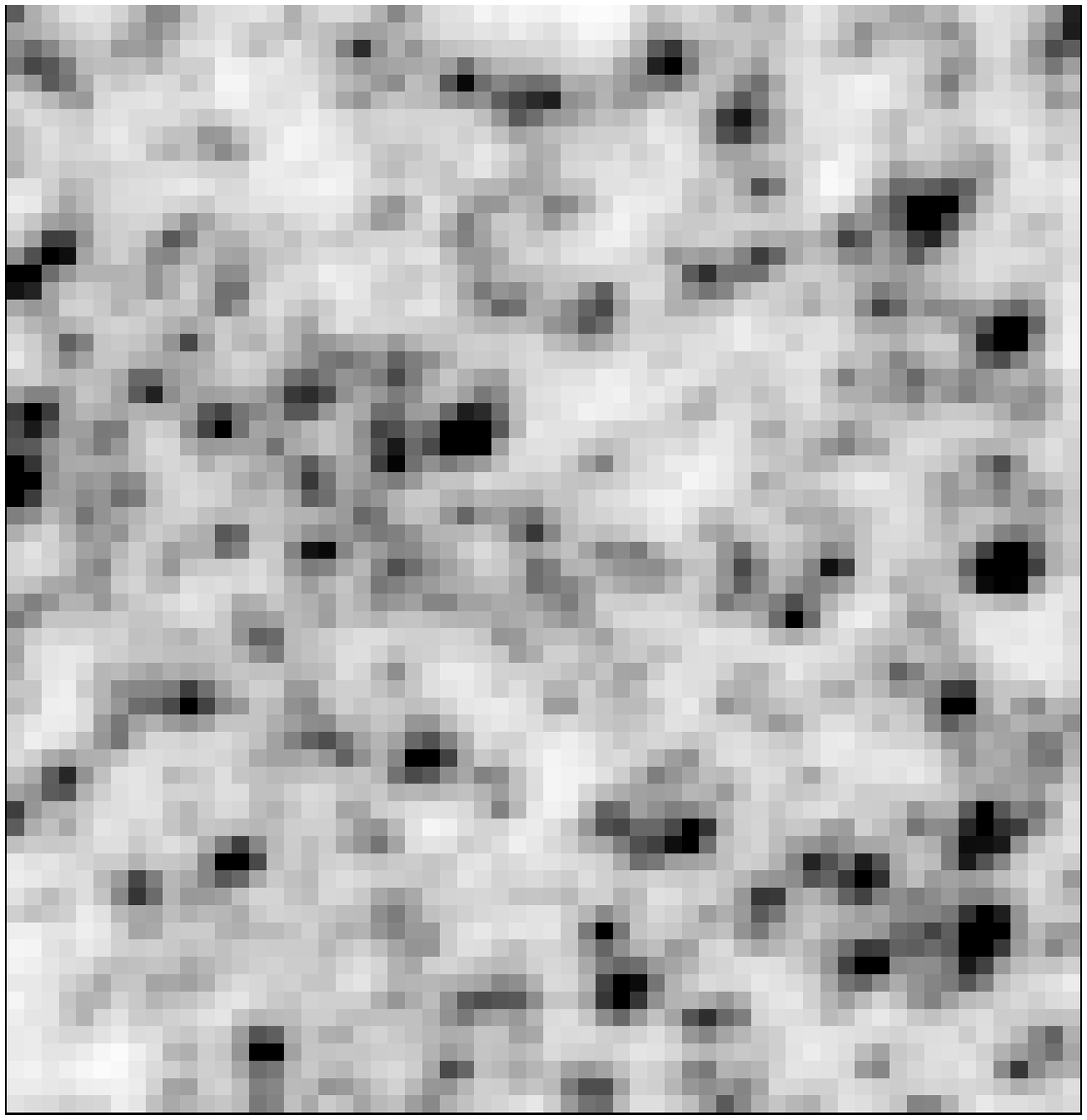}
\includegraphics[width=3in]{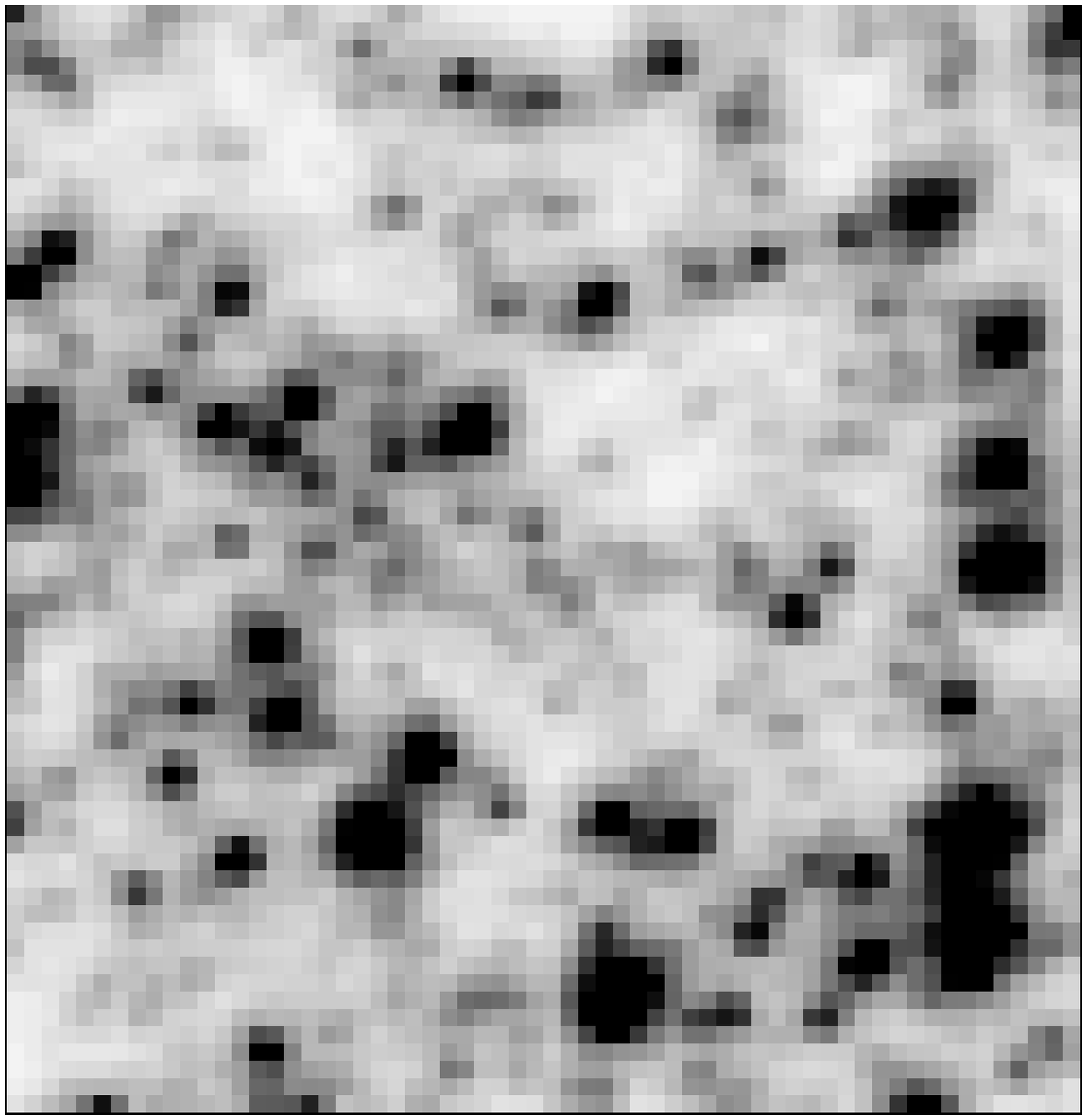}
\end{center}
\figcaption{WFPC2 images of the site of M31~RV, prepared from \HST\/ frames
obtained in 1999 July (11~yr after the outburst) as described in the text. Both
images are $3''\times3''$; north lies $51^\circ$ counterclockwise from up, and
east is $141^\circ$ counterclockwise from up. {\it Left}: $V$-band (F555W) image
prepared by combining 8 images totalling 7200~s exposure time. {\it Right}:
$I$-band (F814W) image prepared by combining 13 images totalling 10,400~s. 
There is no evidence for stars of unusual color at the site of M31~RV, and no
light echo is seen.}
\end{figure}

\begin{figure}
\begin{center}
\includegraphics[width=6in]{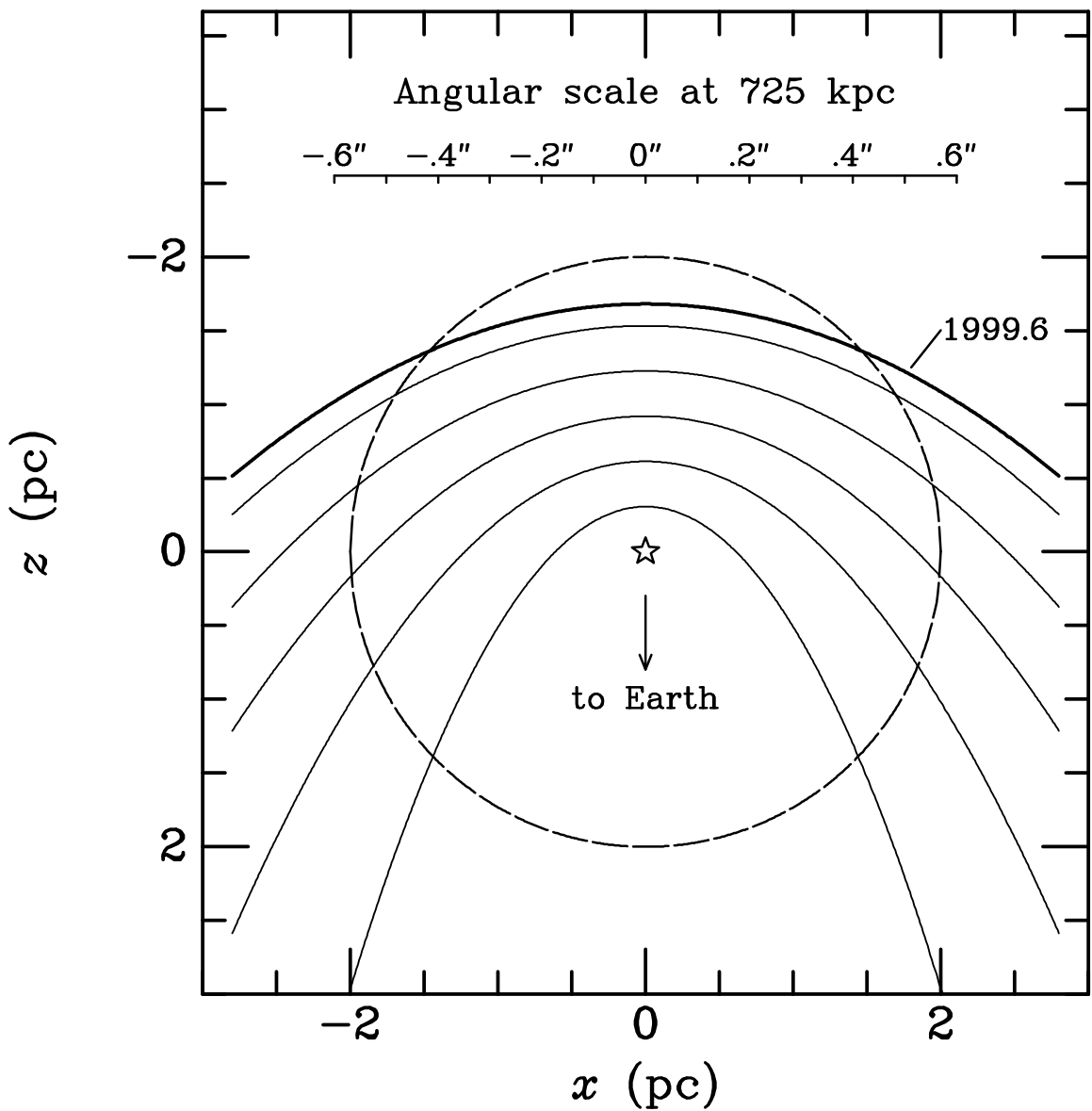}
\end{center}
\figcaption{ Light-echo geometry for M31~RV\null. As explained in the text, at a
time $t$ after the outburst, illuminated dust lies on a paraboloid with the open
end pointing toward the observer. Plotted here are the paraboloids for $t$ = 
2~through 10~yr after the outburst at a spacing of 2~yr (thin lines), and for
the date of the WFPC2 images shown in Fig.~1 ($t=11.0$ yr; thick line labelled
1999.6). Shown at the top is the angular scale at the distance of M31.  The
dashed circle shows a radius of 2~pc around the star, the approximate size of
the circumstellar dust around V838 Mon. If there were similar dust around
M31~RV, a light echo would have been readily visible in Fig.~1.}
\end{figure}

\begin{figure}
\begin{center}
\includegraphics[width=4.2in]{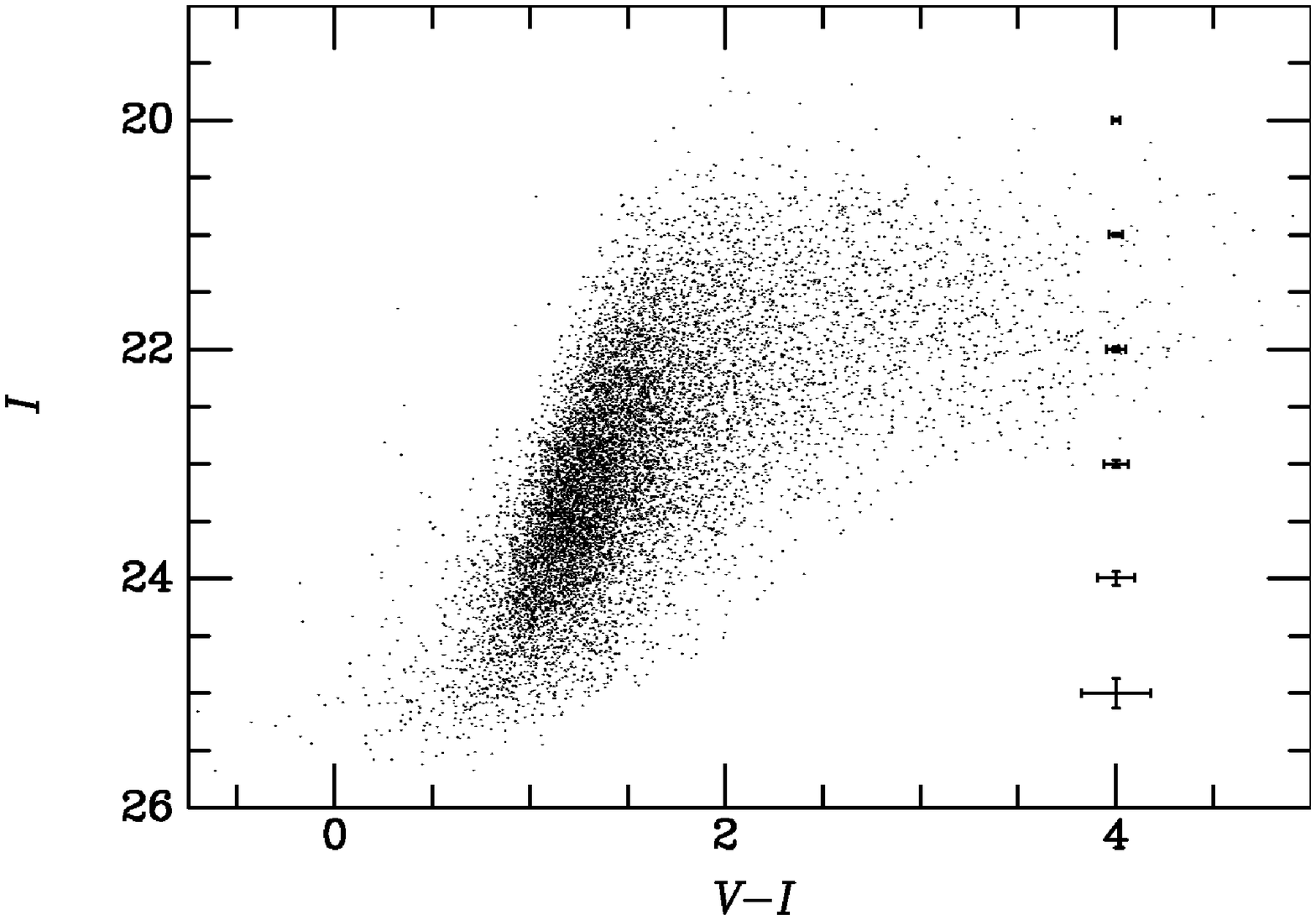}
\end{center}
\smallskip
\begin{center}
\includegraphics[width=4.2in]{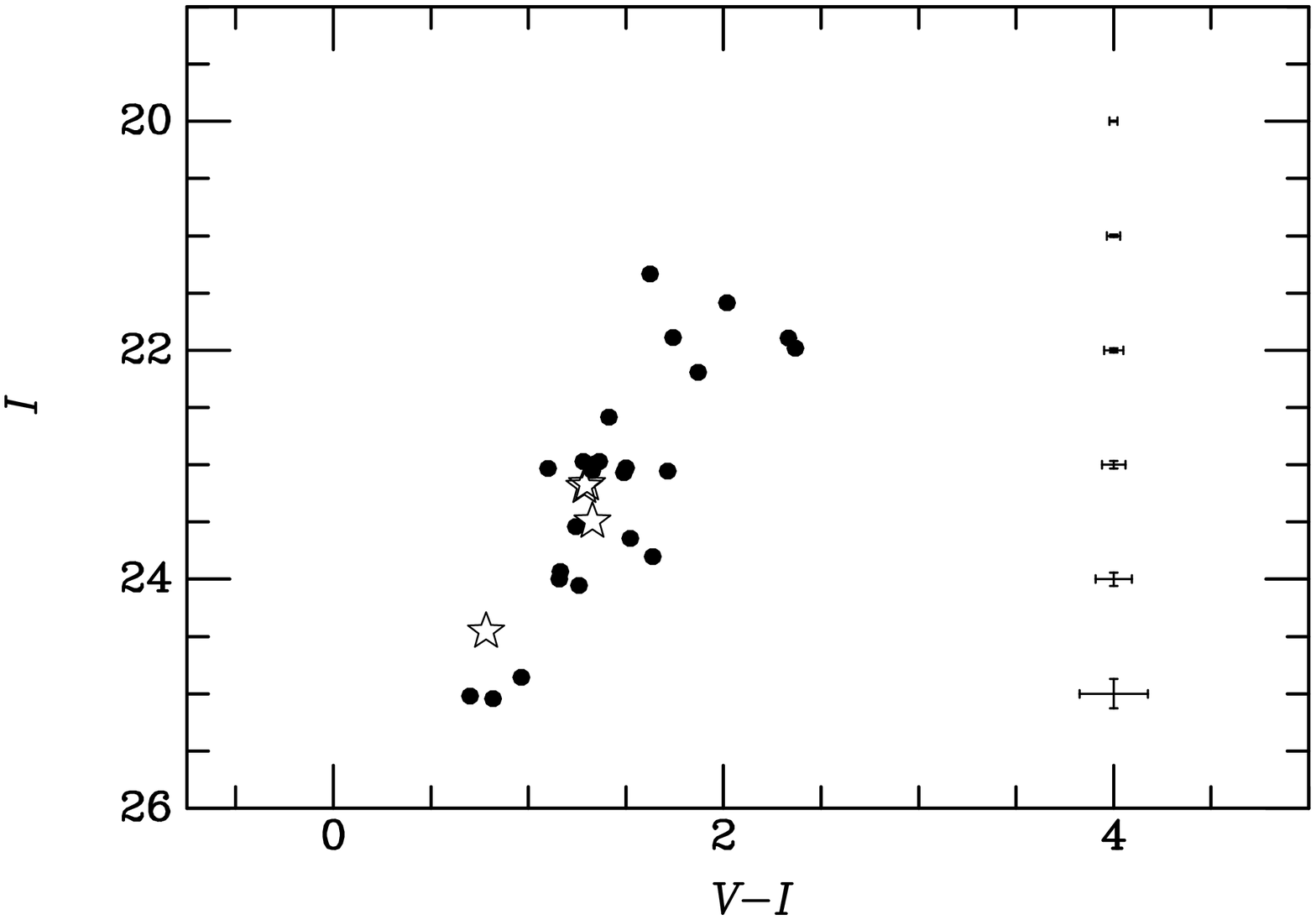}
\end{center}
\figcaption{{\it Top}: {Color-magnitude diagram (CMD) for the field surrounding
the site of M31~RV\null. All of the $\sim$12,000 stars contained in the PC chip
of the WFPC2 camera are plotted. \it Bottom}: CMD for stars lying within a
radius of 6~pixels  ({\it stars}) and within 6 to 18~pixels ({\it filled
circles}) of the site. These radii correspond approximately to
locations within 1$\sigma$ and 3$\sigma$ of M31~RV (see text), and the
corresponding angular radii are $0\farcs27$ and $0\farcs82$. All of the stars
appear to be normal M31 red giants, including all four that are within
$\sim$1$\sigma$ of the M31~RV site. Error bars at the right-hand sides of both
diagrams show the mean photometric errors as functions of $I$ magnitude.}
\end{figure}

\clearpage

\begin{deluxetable}{lll}
\tablewidth{0 pt}
\tablecaption{Absolute Astrometry (Equinox J2000) of M31 RV}
\tablehead{
\colhead{$\alpha$} &
\colhead{$\delta$} &
\colhead{Source}}
\startdata
0:43:02.433 & +41:12:56.17  & This paper \\
0:43:02.42  & +41:12:56.7   & Munari et al.\ 2003, plate 14197 \\
0:43:02.41  & +41:12:57.1   & Munari et al.\ 2003, plate 14222 \\
\enddata
\end{deluxetable}

\begin{deluxetable}{lcccl}
\tablewidth{0 pt}
\tablecaption{Photometry of V4332 Sgr}
\tablehead{
\colhead{Date} &
\colhead{$V$} &
\colhead{$I$} & 
\colhead{$V-I$} & 
\colhead{Source}}
\startdata
1994 Feb (maximum) &   8.5   &	 7.0     & 1.5\phn     & Martini et al.\ 1999 \\    
2003 May 21        & 17.63   &    15.09  & 2.54    & Tylenda et al.\ 2005 \\  
2003 Sep 29        & 17.52   &    15.01  & 2.51    & Banerjee \& Ashok 2004 \\
2004 Jul 16        & $\dots$ &    15.09  & $\dots$ & This paper      \\
2005 Sep 16        & 17.67   &    15.10  & 2.57    & This paper      \\
\enddata
\end{deluxetable}

\end{document}